\documentclass[%
reprint,
floatfix,%
amssymb, amsmath,%
]{revtex4-1}
\usepackage{color}
\usepackage{mathtools}
\usepackage{physics}
\usepackage{esdiff}
\usepackage{subcaption}
\usepackage[utf8]{inputenc}
\def \be {\begin{equation}}
\def \ee {\end{equation}}

\begin{document}
	\title{Investigation of dissipative particle dynamics with colored noise}
		
		\author{Morgane Borreguero $^{(a)}$}
		\email{m.borreguero@tum.de}
		\author{Marco Ellero $^{(b)}$}\author{N.A. Adams $^{(a)}$}	
		\affiliation{(a) Chair of Aerodynamics and Fluid Mechanics \\ Department of Mechanical Engineering, Technical University of Munich}%
		\affiliation{(b) Basque Center for Applied Mathematics, Bilbao, Spain}

		\date{\today}%
		%\revised{August 2010}%
		
		\begin{abstract}
		%\noindent 
		We investigate the behavior of dissipative particle dynamics (DPD) with time-correlated random noise. A new stochastic force for DPD is proposed which consists of a random force whose noise has an algebraic correlation proportional to $1/t$ and is generated by the so called Kangaroo process. We stress the benefits of a time correlated noise in stochastic systems. We show that the system exhibits significantly different properties from classical DPD, driven by Wiener noise. While the probability distribution function of the velocity is Gaussian, the acceleration develops a bi-modal character. Although the fluctuation dissipation theorem may not strictly hold, we demonstrate that the system reaches equilibrium states with fluctuation-dissipation balance. We believe that our explorative research on the DPD model may stimulate the application of modified DPD to unconventional problems beyond molecular modeling.
		%Although one cannot ensure that a fluctuation dissipation theorem is satisfied yet, some strong indications have been highlighted and are shown within this paper. A key result is the fact that a robust steady state is reached by the system. 
		\end{abstract}
	
	\maketitle

\section{Introduction}
%\lettrine[nindent=0em,lines=3]{T}he introduction.
	An important aspect in the choice of a numerical model of a physical problem are the length and time-scale defining the quantities of interest. At macroscopic scales continuum-based methods are appropriate, while Molecular Dynamics (MD) models can capture the microscopic details. A large variety of methods has been proposed for the intermediate mesoscopic scale. E.g. for complex fluids or soft matter mesoscopic scales play an important role. Large time and length scale ranges characterize their behavior and prevent the suitability of a single model to resolve simultaneously different scales. Groot and Warren classified dissipative particle dynamics (DPD) as a mesoscopic simulation method \cite{groot_dissipative_1997}. DPD was introduced as a coarse-grained particle-based stochastic model in a Lagrangian reference frame by Hoogerbrugge and Koelman \cite{hoogerbrugge_simulating_1992}.  Effectiveness has been demonstrated for a wide range of problems, such as multiphase phenomena \cite{pagonabarraga_dissipative_2001}, interaction of polymers, surfactants and water \cite{groot_mesoscopic_2000} and dynamics of membranes \cite{fedosov_multiscale_2010}.
	
	 In this paper we investigate DPD as a numerical model abstracted from a specific application. In \cite{azarnykh_discussions_2018} similarities and differences between the Langevin model and DPD have been highlighted. It has been found that the current auto-correlation functions of the $N$-particle subregime of DPD and the Langevin equations coincide. The Langevin principle \cite{langevin_sur_1908} consists in splitting the motion into two parts: the slowly varying motion, and the rapidly changing properties resolved by a random variable. This separation is appropriate if the characteristic time-scale of the system is larger than the time-correlation of the rapidly varying variables that are modeled by noise. In this case the system is Markovian. The stochastic procedure proposed by Langevin disregards the non-Markovian property of the system. Non-Markovian extensions of the Langevin equation have been applied successfully to many physical systems, e.g. \cite{brustein_langevin_1991,ceriotti_langevin_2009}. 
	 
	Recently, Li et al. \cite{li_incorporation_2015} proposed to incorporate explicit memory effects into DPD via the Mori-Zwanzig formalism. It was found that when there is a lack of time-scale separation, the non-Markovian DPD shows little improvement on the velocity autocorrelation function compared with the standard DPD model. Motivated by this observation, we aim at manipulating memory effects of DPD implicitly. Unlike \cite{li_incorporation_2015}, where the random noise generator is derived according to the second fluctuation dissipation theorem (FDT) from the memory-kernel, we select a specific colored-noise generator and subsequently analyze the friction term together with the FDT. No explicit memory kernel is used here. The Kangaroo process is used as colored-noise generator \cite{peter_hanggi_colored_1995} which has a slowly decaying noise-correlation proportional to $1/t$. In \cite{srokowski_solving_1998} a non-Markovian Langevin equation with a colored noise generated by the Kangaroo process has been proposed. In this case the velocity autocorrelation function (VACF) can be analytically derived, and in the case of the $1/t$ autocorrelation noise, the VACF has an algebraic tail. Also, behavior which is typical for intermittent structures of L\'evy flights has been observed in \cite{srokowski_solving_1998}. We note that physical systems exhibit a slowly decaying correlation are e.g. atomic diffusion through a periodic lattice \cite{igarashi_non-markovian_1988} and ligand migration in biomolecules \cite{hanggi_escape_1986}.  
	
	The paper is structured as follows: In section \ref{Model}, the model is presented in detail together with the theoretical background. Section \ref{ValidationMethod} examines the fluctuation dissipation theorem (FDT) in our context. Numerical results are analized in section \ref{NumericalResults}. The latter also provides comparisons between the standard DPD and the DPD with colored-noise (C-DPD). Finally, we conclude with a discussion, summary of our results and a brief outlook of future research and ongoing work.
	
	%T. Srokowski and M. P\l{}oszajczak \cite{srokowski_solving_1998} proposed to solve a non-Markovian Langevin equation with a colored noise generated by the Kangaroo process.
	
	%"Often the fluctuations represent the cumulative effects of many weakly coupled environmental degrees of freedom" 
	
	%This procedure leads to a lower number of degree of freedom.  
	
	%While simulating complex fluid or soft matter, both MD as well as continuum-based methods have their limitations. Indeed, the presence of a large range of different time and space scales in complex fluid inhibit one to resolve simultaneously all the scales
%\begin{enumerate}[label=$\bullet$]
%	\item Why mesoscale model are interesting
%	\item DPD appropriate for mesoscale \cite{groot_dissipative_1997}
%	\item DPD similar behavior with langevin (Dmitrii) \cite{azarnykh_determination_2016},\cite{azarnykh_discussions_2018}
%	\item Stochastic differential equation - role of noise
%	\item In stochastic differential equation with noise which has correlation time zero the Markovian property get lost (i.e. noise correlation time << system correlation time is only an assumption) \cite{peter_hanggi_colored_1995}
%	\item Advantage/Disadvantage of Markovian-vs-Non-Markovian
%	\item Langevin + colored noise 
%	\item Langevin + Kangaroo (Srokowski) \cite{srokowski_solving_1998}
%	\item Turbulent flows - VACF negative value - non-Gaussian Statistic \cite{li_origin_2005},\cite{zumofen_power_1993}
%	\item What will be investigate in the paper i.e. how it is structured
%\end{enumerate}

\section{The DPD Model}\label{Model}
For standard DPD, the motion of each DPD particle is described by
\begin{equation}\label{DPD1}
\diff{\mathbf{r}_i}{t}=\mathbf{v}_i \text{ , }\diff{\mathbf{v}_i}{t}= \frac{1}{m}\sum_{j\neq i} \mathbf{F}_{ij}.
\end{equation}
The total force acting on particle $i$ is composed of three pairwise interactions
\begin{equation}\label{DPD2}
	\mathbf{F}_{ij} =\mathbf{F}^C_{ij} + \mathbf{F}^D_{ij}+\mathbf{F}^R_{ij} \quad .
\end{equation}
All forces vanish beyond a cutoff radius $r_c$ which we choose to be unity. The conservative force acts along the connecting line between two particles and is soft repulsive
\begin{equation}
	\mathbf{F}_{ij}^{C} = \begin{cases}
	a_{ij}m (1-r_{ij})\hat{\mathbf{r}}_{ij} \quad \text{ if } & r_{ij}<r_c\\
	0 \quad \text{ if } r_{ij}\geq r_c
	\end{cases}\quad .
\end{equation}
In the following we do not consider the conservative force $F^C$. The dissipation and random forces are given by
\begin{subequations}
	\begin{equation}
		F^D_{ij} = - \gamma w^D(r_{ij})(\hat{\mathbf{r}}_{ij}\cdot \mathbf{v}_{ij})\hat{\mathbf{r}}_{ij} \text{ ,}
	\end{equation}
	\begin{equation}
		F^R_{ij} = \sigma w^R(r_{ij})\theta_{ij}\hat{\mathbf{r}}_{ij} \text{ ,}
	\end{equation}
\end{subequations}
where $\mathbf{r}_{ij}=\mathbf{r}_i - \mathbf{r}_j$, $\mathbf{v}_{ij}=\mathbf{v}_i - \mathbf{v}_j$, $r_{ij}=|\mathbf{r}_{ij}|$ and $\hat{\mathbf{r}}_{ij}=\mathbf{r}_{ij}/|\mathbf{r}_{ij}|$. The functions $w^D$ and $w^R$ are weighting functions. Furthermore, $\theta_{ij}$ is a Gaussian random variable with  
\begin{subequations}
	\begin{equation*}
		\langle \theta_{ij}(t)\rangle = 0\text{ ,}
	\end{equation*}
	\begin{equation*}
		\langle\theta_{ij}(t)\theta_{kl}(t') \rangle = (\delta_{ik}\delta_{jl}+\delta_{il}\delta_{jk})\delta(t-t').
	\end{equation*}
\end{subequations}
In order to satisfy the fluctuation dissipation theorem, Espa\~{n}ol and Warren \cite{espanol_statistical_1995} showed that 
\begin{equation}\label{Eq:DPDsigmagamma}
	w^D=(w^R)^2 \text{  and  } \sigma^2 = 2 \gamma k_BT
\end{equation}
must hold. In Eq. \eqref{Eq:DPDsigmagamma} $k_B$ is the Boltzmann constant and $T$ the temperature. We choose the standard kernel function
\begin{equation*}
	w^D = (w^R(r))^2 = \begin{cases}
	    (1-\frac{r}{r_c})^2,& \text{if } r\leq r_c\\
	    0,              & \text{otherwise}
	\end{cases}
\end{equation*}
where $r_c$ is the radius cut-off. The velocity Verlet algorithm is used for time integration.
The equation of motion given by \eqref{DPD1} and \eqref{DPD2} is Markovian since the additive noise $\theta$ is not correlated in time. 

\begin{figure}[h]
	\centering
	\includegraphics[width=0.4\textwidth]{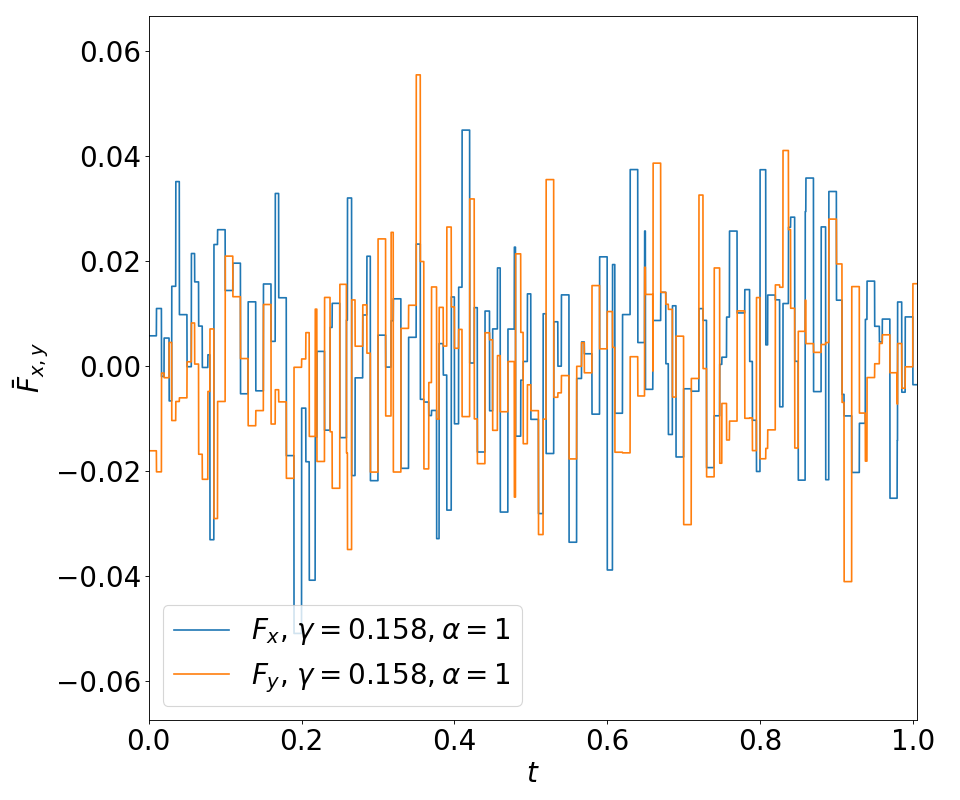}
	\caption{Time evolution of the $x$ and $y$ component of the random force averaged over space.}\label{Fig:FxFy}
\end{figure}

Let us now consider a noise with the following auto-correlation function \cite{srokowski_solving_1998}
\begin{equation}\label{Eq:ACFnoise}
	C_R(t) = \begin{cases}
	\alpha/\varepsilon , & t \leq \varepsilon \\
	\alpha / t, & t > \varepsilon
	\end{cases}\text{ ,}
\end{equation}
for some $\varepsilon>0$. The variable $\alpha$ is the amplitude of the noise. The stochastic process called "the Kangaroo process" (KP) provides such a behavior \cite{brissaud_solving_1974}. Its name originates from its resemblance to L\'evy flights.  The process is defined as a stepwise random function which is constant during a time interval $\Delta t$ which is a random number. Although the Kangaroo process can be formulated for higher dimensions, here we focus on two dimensional systems. The resulting stochastic force for particle $i$ is given by
%Let $m$ be a constant that the function may reach for some finite and maybe repeated time; the frequency of jumping times $\nu(m)$ is function of the value of the process itself.
\begin{equation}\label{Eq:mximyi}
	\mathbf{F}^R_i(t) =  \begin{pmatrix} m_{x,i} \\ m_{y,i} \end{pmatrix} = \begin{pmatrix}
	\sqrt{\alpha} \cos (\phi_i/\sqrt{\varepsilon})) \\ \sqrt{\alpha} \sin (\phi_i/\sqrt{\varepsilon})
	\end{pmatrix}\text{ ,}	
\end{equation}
where $\phi_i$ is a random number uniformly distributed in the interval $(0,1)$. The force $F^R_i$ stays constant during an interval $\Delta t$ which is defined as
\begin{equation}
	\Delta t = \varepsilon / \phi_i .
\end{equation}
If $t>t_n+\Delta t$ a new $\phi_i$ is generated as well as a new time interval within which the process stays unchanged, and so on. In Fig. \eqref{Fig:FxFy} the force distribution is shown.

The random force \eqref{Eq:mximyi} is inserted into the equation of motion
\begin{equation}\label{Eq:EquationOfMotion}
	m\diff{\mathbf{v}_i}{t}=\sum_{j\neq i} \mathbf{F}^D_{ij} + \mathbf{F}^R_i
\end{equation}
and gives the C-DPD version of DPD.
 Assuming the system to be in equilibrium with a heat bath of temperature $T$, it has to satisfy the fluctuation dissipation theorem for this given $T$ \cite{prigogine_calculation_2007}. The fluctuations must be linked to the dissipation in such a way that no loss or gain of energy occurs in the system. The temperature has to fluctuate around a constant. It is for many reasons not straightforward to fulfill this condition. Firstly, the forces on the particle $i$ are of different kind, one is a pairwise interaction while the other is a background random fluctuation. Secondly, the stochastic force does not vanish for $r>r_c$. The dissipative force is defined through a kernel and has finite support.
 
Since the stochastic force is time correlated a discussion on the necessity of a memory kernel as in \cite{li_incorporation_2015} is in order. One standard procedure would be to derive the Fokker-Planck equation and solve it for its steady state $\rho^{eq}$ \cite{marsh_theoretical_nodate,espanol_statistical_1995}. The Fokker-Planck equation is a special case of the forward Chapman-Kolmogorov equation (CKE)

\begin{align*}
\frac{\partial}{\partial t}\rho(\mathbf{r},t) =& -\frac{\partial}{\partial \mathbf{r}}\left[A(\mathbf{r},t)\rho(\mathbf{r},t)\right]+\frac{\partial^2}{\partial \mathbf{r}^2}\left[B(\mathbf{r},t)\rho(\mathbf{r},t)\right]\\&+\int d\mathbf{r}' \left[W(\mathbf{r}\vert \mathbf{r}',t)\rho(\mathbf{r}',t)-W(\mathbf{r}'\vert \mathbf{r},t)\rho(\mathbf{r},t)\right]
\end{align*}
 for which the integral term vanishes. For processes with jumps we have to determine $A$, $B$ and 
 \begin{equation}
 	W(\mathbf{r}\vert \mathbf{r}',t)=\lim\limits_{\Delta t\to 0}p(\mathbf{r},t+\Delta t\vert \mathbf{r}',t)/\Delta t \quad ,
 \end{equation}
 where $p$ is a transition probability defined by the nature of the chosen random process. According to \cite{kaminska_solving_2003}, for KP functions $A$ and $B$ vanish. The CKE for DPD with colored noise generated by the KP can be superimposed by the functions $A$ and $B$ for Eq. \eqref{Eq:EquationOfMotion} without stochastic force and by the jump integral for the KP random term
 \begin{align*}
 	\frac{\partial}{\partial t}\rho(\mathbf{r},t)=& -\left[ \sum_i \frac{\mathbf{p_i}}{m}\frac{\partial }{\partial \mathbf{r}_i}\right]\rho(\mathbf{r},t)\\
 	&+\sum_i \pdv{\mathbf{p}_i}\left[\sum_{j\neq i}\mathbf{e}_{ij}\gamma w^D (\mathbf{e}_{ij}\cdot \mathbf{v}_{ij})\right]\rho(\mathbf{r},t)\\
 	&-\nu(\mathbf{r})\rho(\mathbf{r},t)+Q(\mathbf{r})\int \nu(\mathbf{r}')\rho(\mathbf{r}',t)d\mathbf{r}'
 \end{align*}
 where $\nu$ is the jump frequency and $Q$ the size of a jump.
 In order to address the question whether a FDT can be satisfied we assume that the dissipative term in Eq. \eqref{Eq:EquationOfMotion} is convolved with a memory kernel $K$. Eq. \eqref{Eq:EquationOfMotion} takes the form of the generalized Langevin equation
 \begin{equation}
 		m\diff{\mathbf{v}_i(t)}{t}=-\int_0^tK(t-s)\sum_{j\neq i} \mathbf{v}_{i}(s)ds + \mathbf{F}^R_i(t).
 \end{equation}
 From the latter equation one can see that the FDT is satisfied if
 \begin{equation}
 	C_R(s) = k_BTK(t-s)\quad .
 \end{equation}
 As approximation, the memory kernel $K$ can be neglected beyond a time cut off \cite{li_incorporation_2015}, i.e. the history of the system is taken into account only for $t \leq Ndt$, where $dt$ is the time step of the integration-scheme. So, if $t\leq Ndt$ the shape of the memory kernel follows from the noise auto-correlation function, otherwise it vanishes. However for the Kangaroo process, upon choosing 
 \begin{equation}\label{Eq:AssumptionMemoryKernelNeglectable}
 \varepsilon = Ndt
 \end{equation}
 the random force produces a constant memory kernel, see \cite{li_incorporation_2015}, Eq. (16), for $0< t \leq \varepsilon$, and time can be collapsed into a recalibrated DPD dissipation term. 

\section{The fluctuation dissipation theorem}\label{ValidationMethod}
	\begin{table}
	\caption{\label{tab:DPDSolvent}Input parameters of the C-DPD and DPD solvent. }
	\begin{ruledtabular}
		\begin{tabular}{ll}
			Domain size ($V = L_x\times L_y$) & $60 \times 60$\\
			Total number of particles ($N_p$) & $42^2$\\ 
			Mass (m) & $1$\\
			Temperature ($k_B T$)  &$1$\\
			Time step ($dt$) & $0.001$\\
			Density ($\rho$) & $N_p/V$\\
			Simulation length in time step ($N_t$) & $10 ^{6}$\\
			Cut-off radius ($r_c$) & $1$\\
			\textbf{C-DPD } & \\
			\hline
			Dissipative coefficient ($\gamma$)    &$0.158$\\
			Stochastic coefficient ($\alpha$) & $1$\\
			Maximal non-correlated time ($\varepsilon$) & $0.05$\\
			\textbf{DPD}\\
			\hline
			Dissipative coefficient ($\gamma$)    &$0.5$\\
			Stochastic coefficient ($\sigma$) & $2\gamma k_B T$\\
		\end{tabular}
	\end{ruledtabular}
\end{table}

 Unless specified differently, simulations presented in this paper have been performed with parameters presented in Table \ref{tab:DPDSolvent}. Each particle is initialized with a random velocity.%the following set of parameters: $dt =0.001$, $\varepsilon=0.05$, $\rho_{ref}=4$, the 2D box has length $L_x=60$ and height $L_y=60$. The system contains $42\cdot 42$ particles.
 
 The fluctuation dissipation theorem (FDT) is a crucial element in determining the prevailing thermodynamic regime. When it applies, the system temperature coincides with the actual physical temperature. For this reason it is necessary to check whether a FDT is satisfied or not \cite{espanol_statistical_1995}. In cases where an analytic derivation is difficult, one may verify whether the conditions for the FDT are satisfied by simulations. First, the system should reach a stationary state, which is a necessary condition. Furthermore, the stationary state has to be unique. The system has to fluctuate around a constant temperature, independently of the chosen initial condition.
 
\begin{figure}[h]
	\centering
  	\includegraphics[width=0.45\textwidth]{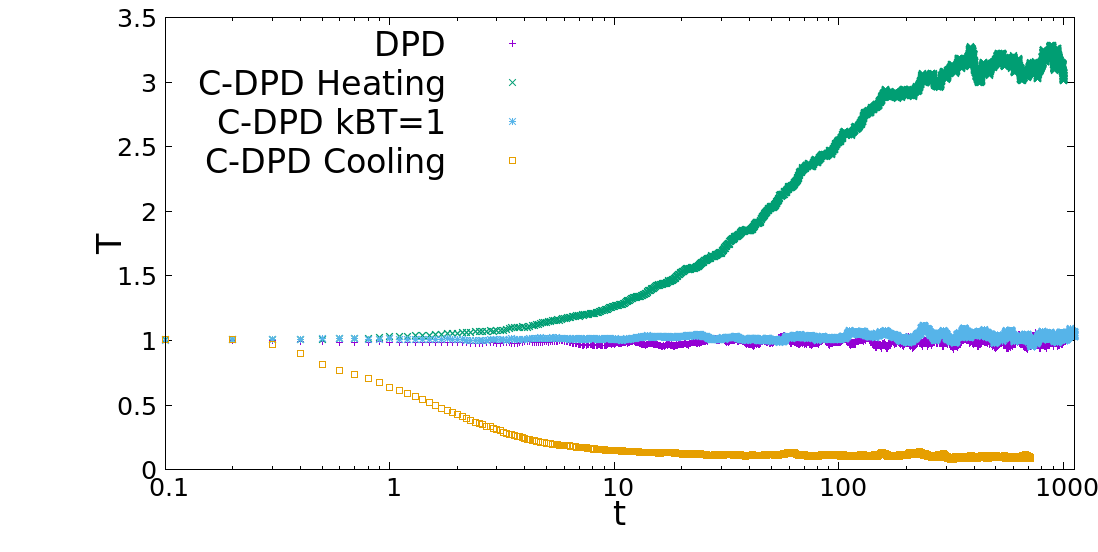}
  	\caption{Time evolution of the mean temperature over all particles of the system. See Table \ref{tab:RealizationSummary}.}\label{Fig:T_t}
\end{figure}

For $\alpha=1$, different friction coefficients $\gamma$ are tested. The results are compared with a DPD simulation with $\gamma_{DPD} = 0.5$. The first observation is that for a specific choice of $\gamma$, the temperature of C-DPD fluctuates around the same equilibrium value as that of DPD, see Fig. \eqref{Fig:T_t}. When the stochastic force dominates the dissipation of the system, the system heats up and reaches a constant temperature after some transient. When the dissipation force overwhelms the effects of the stochastic term, the system cools down. In Fig. \eqref{Fig:T_t}, for the yellow set of data points, the temperature fluctuates around $T=0.34$ but remains stationary after $t\approx 40 $, which implies that the system has cooled to its equilibrium temperature. For the green data set the temperature grows until it also reaches an equilibrium value. These observations are further discussed in sections \ref{NumericalResults} and \ref{SummaryAndDiscussions}. 

\begin{figure}
	\centering
	\includegraphics[width=.8\linewidth]{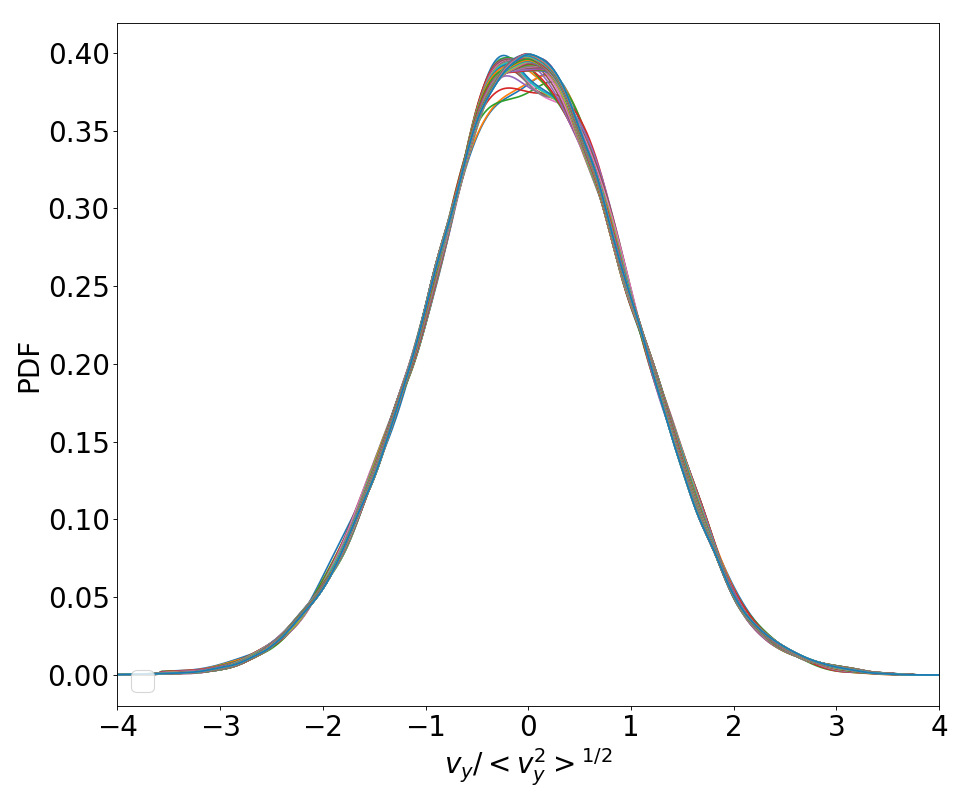}
	\caption{Time series of the pdf of the of the $y$ component of the velocity.}
	\label{Fig:Timeseries}
\end{figure}

\begin{figure}[h]
 	\centering
 	\includegraphics[width=0.4\textwidth]{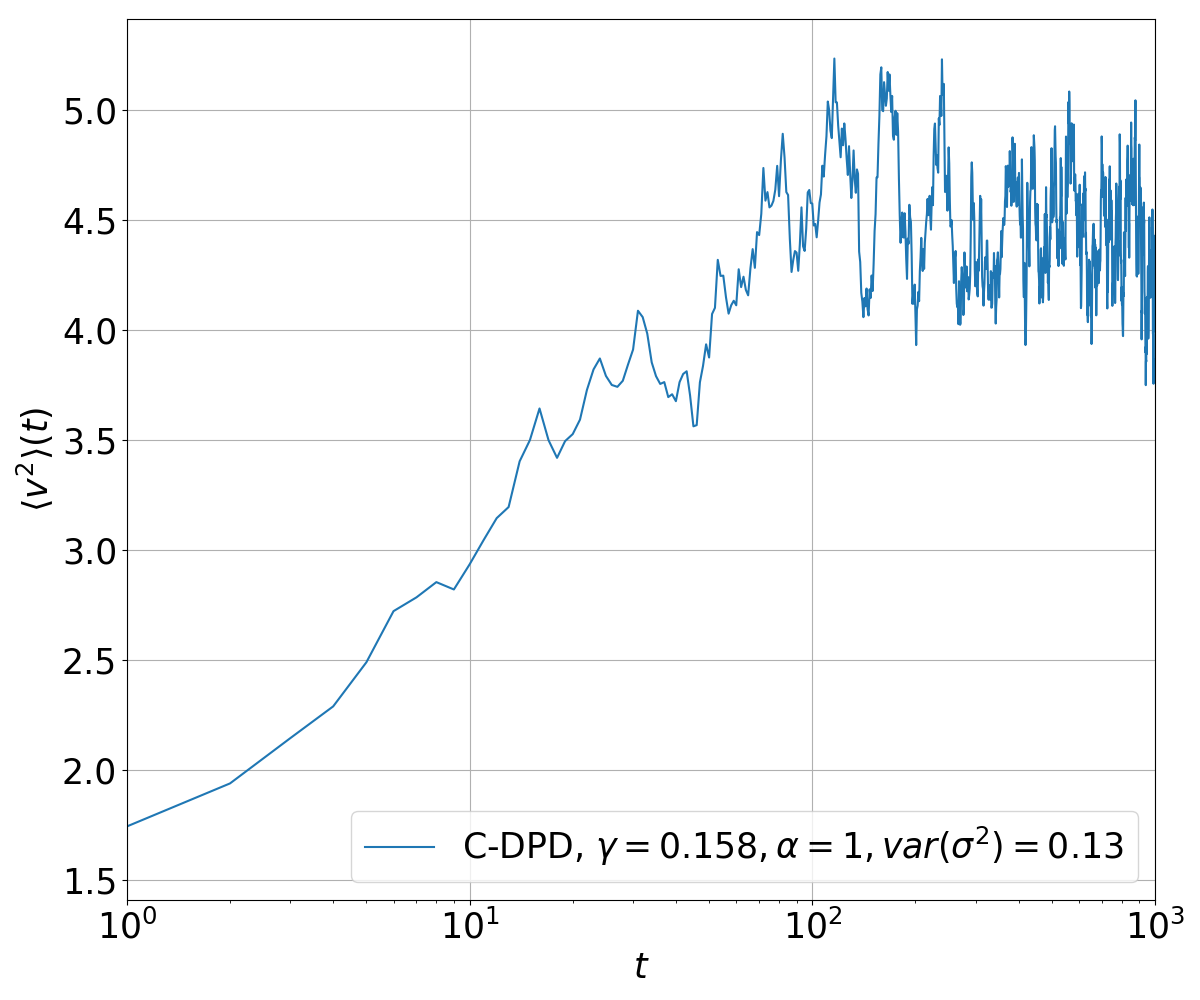}
 	\caption{Variance of the squared velocity.}\label{Fig:VarV2_t}
 \end{figure}

In Fig. \eqref{Fig:Timeseries} the probability density function (pdf) of the velocity in the $y$-direction at different times is represented. It appears that apart from small deviations due to the finite sample size (finite number of particles) the pdf is stationary since it does not change with time. Furthermore, in Fig. \eqref{Fig:VarV2_t} the variance of the squared velocity is plotted in a lin-log scale which shows that a stationary state is reached. Also the variance of the latter quantity stays in the same range as the one obtained from standard DPD.
 
 % Using a fitting procedure, we find for each $\alpha$ its corresponding friction coefficient.  

\begin{figure}[h]
	\centering
	\includegraphics[width=0.5\textwidth]{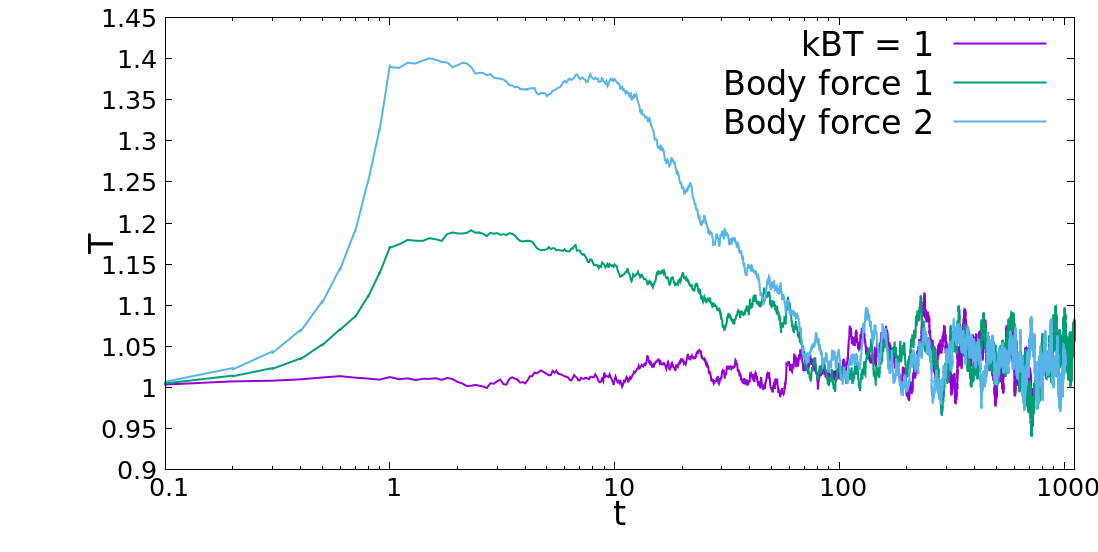}
	\caption{Time evolution of the mean temperature. The violet curve has been produced with an random initial velocity. }\label{Fig:T_iterations}
\end{figure}

In the following we address the issue whether the C-DPD system reaches a unique stationary state, independently of the initial state. Three different simulations are performed, all with the same coefficients for the dissipative and random forces. For each one different initial data are chosen. In Fig. \eqref{Fig:T_iterations} the temperature evolution for these three data sets are plotted. The data "Body force 1" corresponds to a pre-run with a body force $F_x = \sin (2 \pi y /L_y)$, data "Body force 2" corresponds to body force $F_x = 3/2 \sin (2 \pi y /L_y)$ each applied until $t = 1$ and turned off afterwards. See Table \ref{tab:RealizationSummary}. The first case, without body force, corresponds to a simulation for which the initial velocity has been chosen randomly but no body force is applied to the flow. For the other two cases, with body force 1 and 2, for the first 1000 iterations an external body force $F_x$ is applied and then turned off. We choose $F_x$ to be sinusoidal
\begin{equation}
	F_x = \sigma \sin (2 \pi y/L_y) \quad .
\end{equation}
We see that independently of the initial state, the system relaxes towards the same stationary state. The pdf of the velocity and the acceleration of these three examples do not differ from each other.

\begin{figure}[h]
	\centering
	\includegraphics[width=0.4\textwidth]{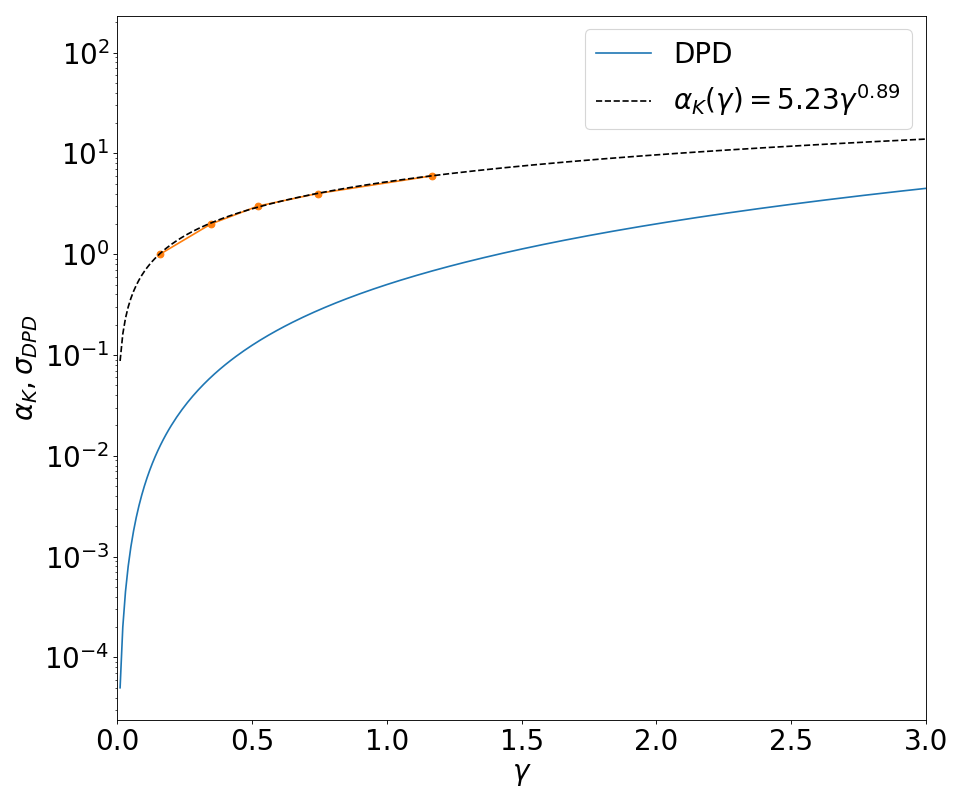}
	\caption{ The fluctuation-dissipation relation for C-DPD. The power law for the blue curve is given by Eq. \eqref{Eq:DPDsigmagamma}.}\label{Fig:Sigma_gammaRelation}
\end{figure}

We here performed many realizations with different stochastic coefficients $\alpha=1,2,3,6$. For each value of $\alpha$ one can find a friction coefficient for which the system reaches a stationary state. By iterating this process and a fitting procedure, we calibrate a $\alpha(\gamma)$ function, Eq. \eqref{Eq:FluctuationDissipationRelation}. Empirically we have established that a FDT may be satisfied for the relation

 %In Fig. \eqref{Fig:VarVy_gamma} the PDF of the $y-$component of the velocity is represented and one can easily seen that the results are in good agreement with the expectations. 
 % \begin{figure}[h]
 % 	\centering
 % 	\includegraphics[width=0.4\textwidth]{./VarVy_gamma.png}
 % 	\caption{The variance of the velocity in $y$-direction versus $\gamma$. (recall that since the system is symetric it does not make any difference if we take the $x$ or the $y$ component). The straight line in log-log scale is observed as expected.}\label{Fig:VarVy_gamma}
 % \end{figure}
 
 %Finally, the set of parameter $(\alpha,\gamma)$ that satisfied these requirements are plotted on a graphic so that their dependencies can be read out of it. For DPD we have \eqref{Eq:DPDsigmagamma} while from the DPD with Kangaroo process it came out from simulations and Fig. \eqref{Fig:Sigma_gammaRelation}
 \begin{equation}\label{Eq:FluctuationDissipationRelation}
 \alpha = 5.23\gamma^{0.89}.
 \end{equation}

%All those results confirm that the assumptions made by choosing $\varepsilon$ following relation \eqref{Eq:AssumptionMemoryKernelNeglectable} is reasonable and that a FDT is fulfilled. The system fluctuates around the given temperature and reaches a steady-state.
%\begin{enumerate}[label=$\bullet$]
%	\item Explain the set up
%	\item Stochastic force over time Fig. (1)
%	\item Trajectory Fig. (2), the first 25 time step, less change of direction with kangaroo
%	\item Fig. (3) the temperature should fluctuate around a constant value, namely the chosen value. For $T=1$. 
%	\item Fig. (4) Comparison with DPD, the variance of the velocity should not change. As it is the case in DPD.
%	\item By a fitting process: Fig. (5) find the relation between $\alpha$ and $\gamma$.
%	\item to check process: Fig. (6), (7),(8)
%\end{enumerate}

	\begin{table}
	\caption{\label{tab:RealizationSummary}C-DPD realization's summary }
	\begin{ruledtabular}
		\begin{tabular}{lll}
			\textbf{Name } &$(\gamma, \alpha)$  & initial condition\\
			\hline
			$k_B T = 1$     &$(0.158,1)$ & Random velocity\\
			Cooling  & $(10,1)$ & Random velocity\\
			Heating & $(0.05,1)$ & Random velocity\\
			\hline
			Body force 1 & $(0.158,1)$ & $F_x = \sin (2 \pi y /L_y)$ for $t\leq 100$\\
			Body force 2 & $(0.158,1)$ & $F_x = 1.5 \sin (2 \pi y /L_y)$ for $t\leq 100$\\
		\end{tabular}
	\end{ruledtabular}
\end{table}

\section{Discussion of results}\label{NumericalResults}
\begin{widetext}
\begin{figure}[h]
	\centering
	\begin{subfigure}{.5\textwidth}
		\centering
		\includegraphics[width=.7\linewidth]{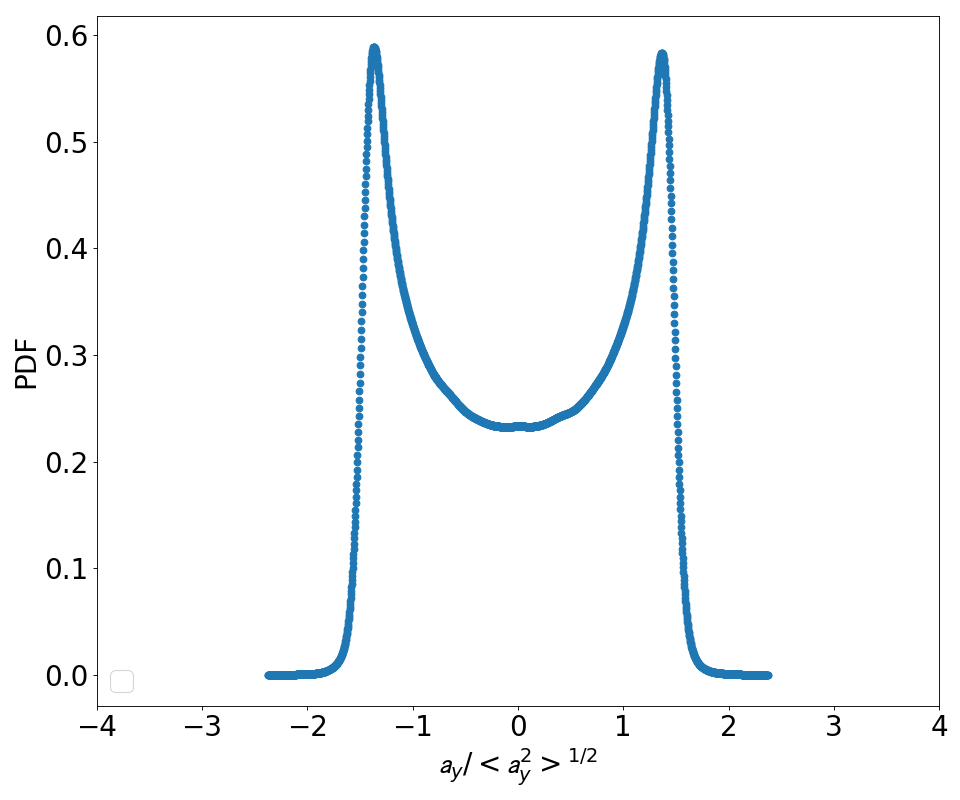}
		\caption{Acceleration pdf}
		\label{fig:sub1}
	\end{subfigure}%
	\begin{subfigure}{.5\textwidth}
		\centering
		\includegraphics[width=.7\linewidth]{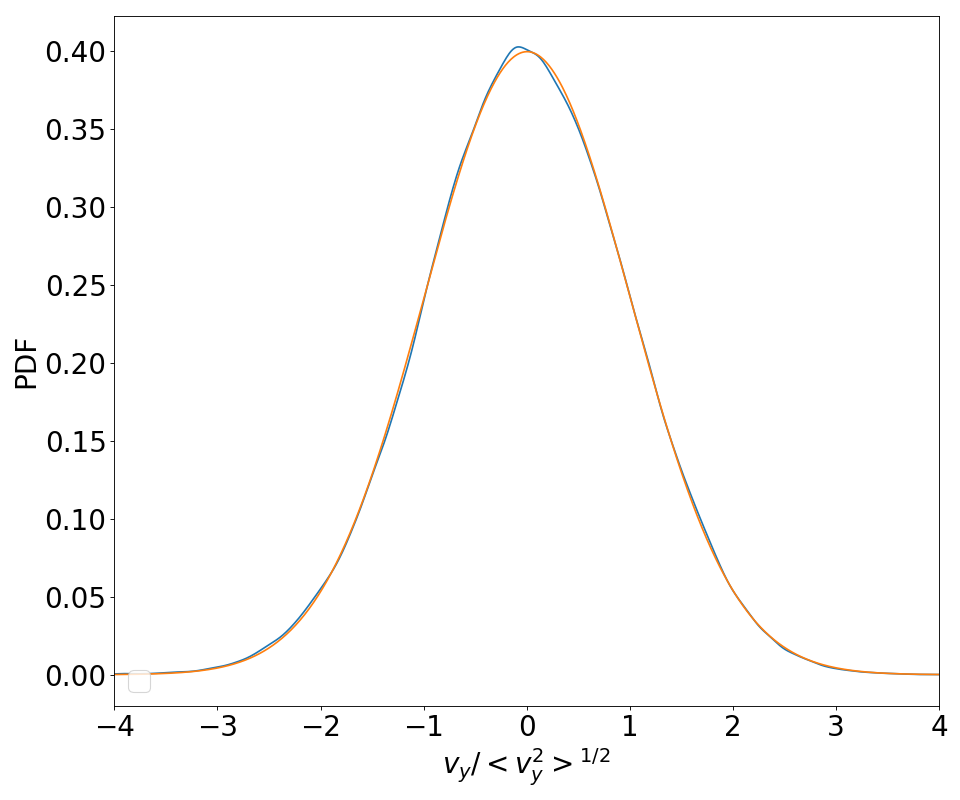}
		\caption{Stationary velocity pdf, orange curve is Gaussian fit.}
		\label{fig:sub2}
	\end{subfigure}
	\caption{Stationary pdf of the $y$ component of the acceleration normalized by its standard deviation. }
	\label{Fig:PDF_F}
\end{figure}
\end{widetext}

Results presented in this section have been produced using mostly the empirical fluctuation-dissipation relation \eqref{Eq:FluctuationDissipationRelation}.
In Fig. \eqref{fig:sub2}, a Gaussian (orange curve) has been used as fit to indicate that the pdf of velocity is Gaussian. While for DPD the pdf of the acceleration also follows a Gaussian, for C-DPD it results in a bimodal Gaussian. The realization used to generate Fig. \eqref{Fig:PDF_F} correspond to $k_BT =1$ in Table \ref{tab:RealizationSummary}. The system has two different most likely states. %Up to very small variations, the two Gaussians have the same variance. %The mean value of the Gaussian are equal with opposite sign, i.e. $\mu = \pm \sqrt{\alpha}$. 
 
\begin{figure}[h]
	\centering
	\includegraphics[width=0.4\textwidth]{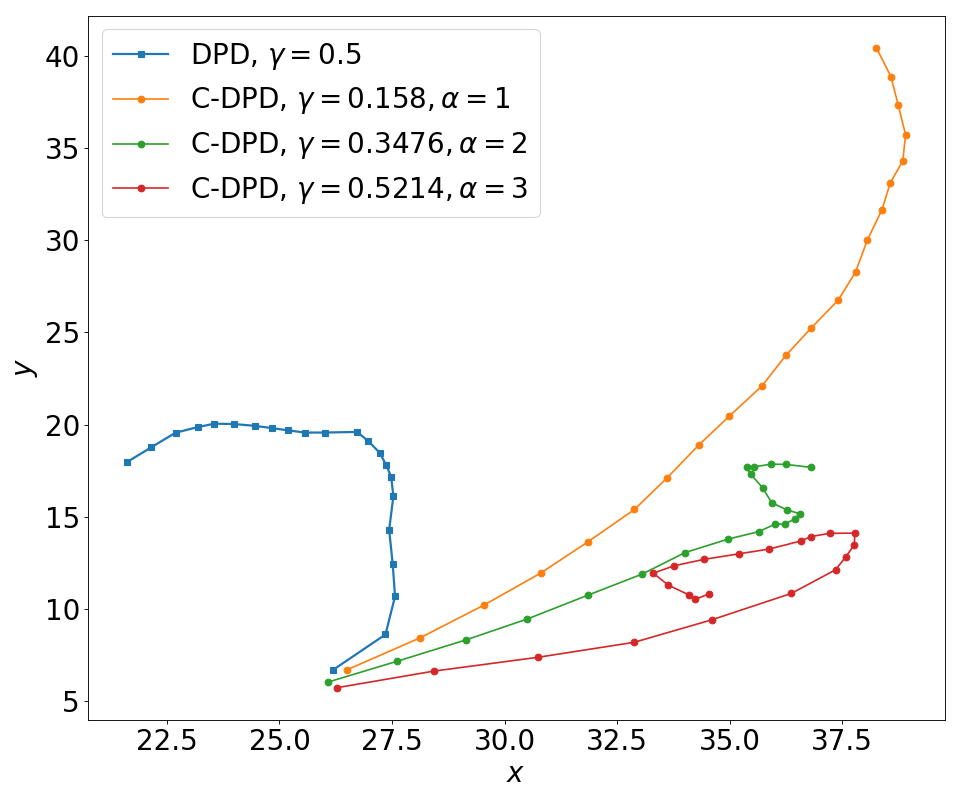}
	\caption{Particle trajectories for $20$ time steps.}\label{Fig:Trajectories}
\end{figure}

 This kind of behavior may be explained by the coexistence of two different populations at stationary state. They are categorized by their acceleration. A particle may reside in one population for a finite time (which depends on the parameter $\varepsilon$) before it jumps to the other. One expects trajectories with short segment of free flight before a sudden change of direction, see Fig. \eqref{Fig:Trajectories}, where trajectories for the first $20$ time steps are shown. Note that the notion of free flight is used in a descriptive sense. During such free flights, the particle undergoes constant acceleration. While for the orange trajectory no change in direction yet has happened, the red as well as the green trajectories clearly show such events. Also in comparison to the DPD, for the three C-DPD simulations free flights are observed.

%\begin{figure}[h]
%	\centering
%	\includegraphics[width=0.4\textwidth]{./PDF_F2.png}
%	\caption{Probability density function of the acceleration in the $y$-direction. Note that the system is completely symmetric so that it does not make a difference whether we look at the $x$ or $y$ component. The two overlapped Gaussian distribution for the C-DPD are well seen. Up to very small variations, the two Gaussian have the same variance. }\label{Fig:PDF_F}
%\end{figure}

%An interesting observation is made by looking at Fig. \eqref{Fig:VarF_gamma}. Even though, the nature of the PDF of the acceleration of the C-DPD diverges from the one of DPD, the trend remains. The variance of the acceleration in $y$-direction (recall that $m=1$) grows with a very similar slope as $\gamma$ increases. \textcolor{red}{TODO: compute the slope \& maybe develop here discussion \& why this shift?}. 
%\begin{figure}[h]
%	\centering
%	\includegraphics[width=0.4\textwidth]{./VarF_gamma.png}
%	\caption{\textcolor{red}{TODO}}\label{Fig:VarF_gamma}
%\end{figure}

\begin{figure}[h]
	\centering
	\includegraphics[width=0.4\textwidth]{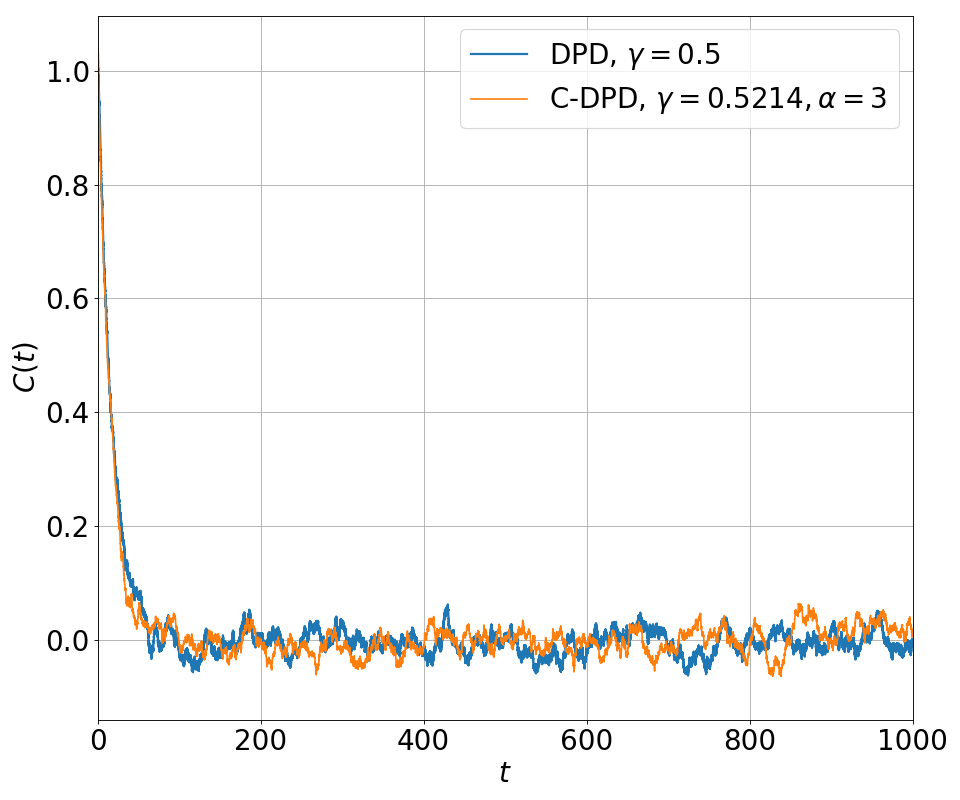}
	\caption{Velocity auto-correlation function.% We choose a realization for C-DPD for which $\gamma_{DPD}\approx \gamma_{C-DPD}$.
	}\label{Fig:VACF}
\end{figure}

The velocity auto-correlation function (VACF) exhibits a power law such as Eq. \eqref{Eq:VACFfit} for the DPD as well as for the C-DPD, Fig. \eqref{Fig:VACF}. The VACF is connected to the mean free path \cite{zumofen_power_1993}. The C-DPD and the DPD VACF are very similar. Two regimes can be distinguished, first an exponential decay and at later times an algebraic dependence. The VACF decays rapidly towards their stationary value $C_{\infty}$. For both models $C_{\infty}$ is almost zero. No significant negative tail can be observed, which indicates essentially classical diffusion. 

The diffusion coefficient can be computed by means of the VACF
\begin{equation}\label{Eq:DissipationCoefficient}
\mathcal{D} = \int_{0}^{\infty} C(t)dt\quad .
\end{equation}
\begin{figure}[h]
	\centering
	\includegraphics[width=0.4\textwidth]{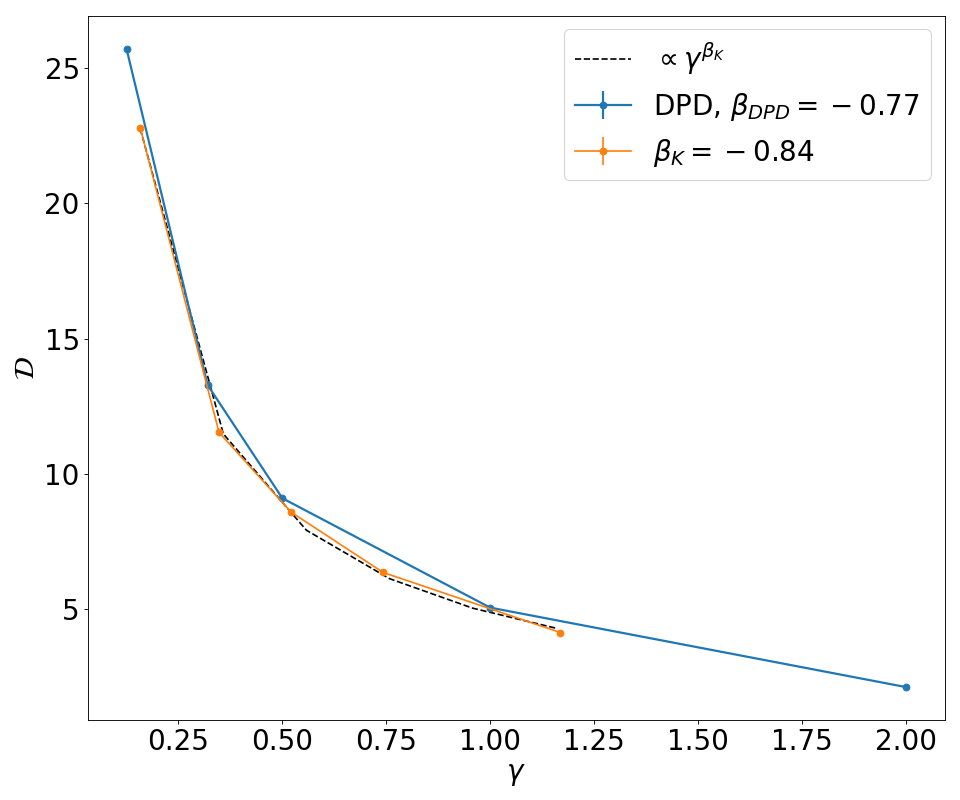}
	\caption{Diffusion coefficient versus friction coefficient.}\label{Fig:Dissipation_gamma}
\end{figure}

The results in Fig. \eqref{Fig:Dissipation_gamma} show that the relation between the dissipation coefficient and the friction coefficient $\gamma$ for both versions of DPD are very similar.

Referring to the issue of the cooling and heating behavior, Fig. \eqref{Fig:T_t}, we observe that for realizations where the dissipation overwhelms the fluctuation the equilibrium temperature of the system is smaller than the input temperature. However, it assumes a stationary value after a transient. When the dissipation is almost negligible, the temperature grows, but the system again reaches a stationary temperature value. 

\begin{figure}[h]
	\centering
	\includegraphics[width=0.4\textwidth]{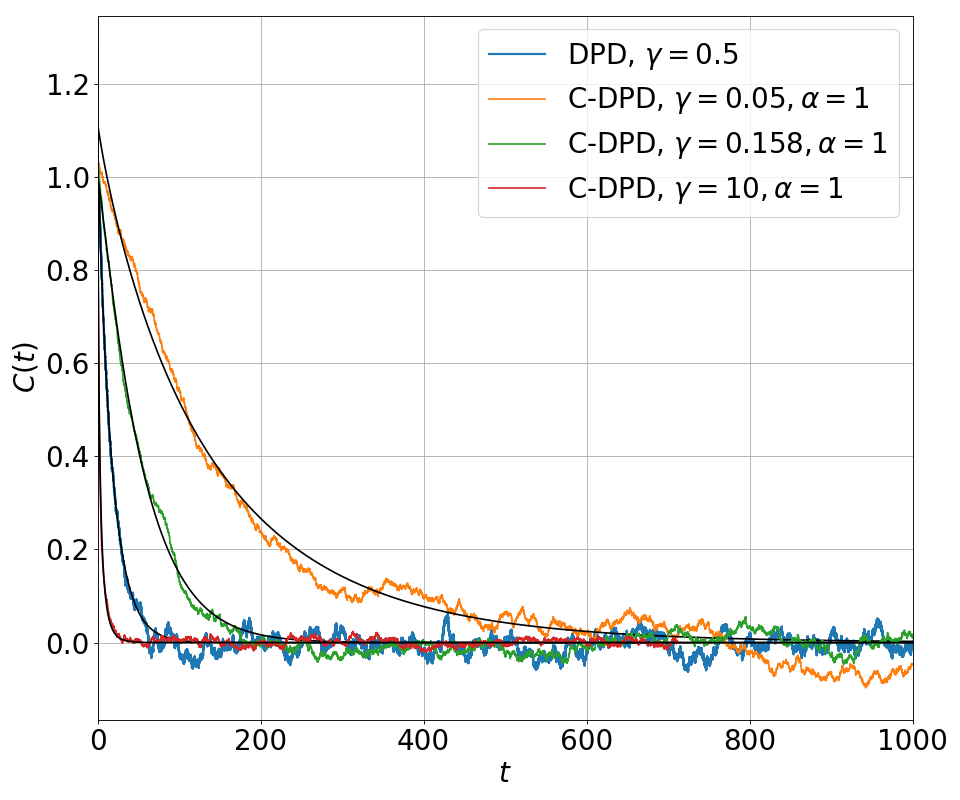}
	\caption{Velocity autocorrelation function fitted by function \eqref{Eq:VACFfit}.}\label{Fig:VACF_alpha_1_DPD_0.5}
\end{figure}

\begin{figure}[h]
	\centering
	\includegraphics[width=0.4\textwidth]{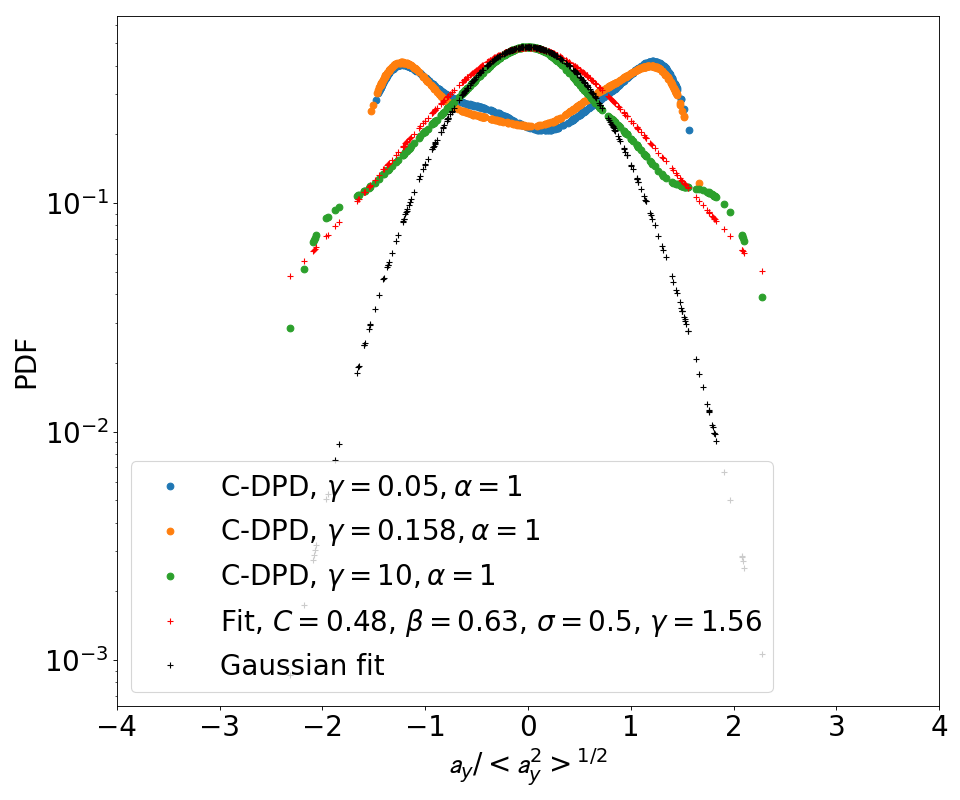}%PDF_F_alpha_1_DPD_05.png}
	\caption{Probability distribution function of the $y$-component of the particle acceleration normalized by its standard deviation for different parameters $\gamma$.}\label{Fig:PDF_F_alpha_1_DPD_0.5}
\end{figure}

We investigate both cases where the parameter $\gamma$ is either too big or too small to satisfy relation \eqref{Eq:FluctuationDissipationRelation}. In Fig. \eqref{Fig:VACF_alpha_1_DPD_0.5} the VACF for these cases are plotted for constant $\alpha$. The black lines correspond to the fit
\begin{equation}\label{Eq:VACFfit}
g(t) = a e^{-(bt)^c}.
\end{equation}
The calibrated parameters $c$ for the corresponding realization are presented in table \ref{tab:journals}. 
	\begin{table}
		\caption{\label{tab:journals}Decay of the VACF}
		\begin{ruledtabular}
			\begin{tabular}{ll}
				\textbf{C-DPD $\gamma$} & \textbf{$c$} \\
				\hline
				$0.05$ & $0.90$\\
				$0.158$  &$0.70$\\
				$10$    &$0.62$\\
				\\
				\textbf{DPD $\gamma$} & $c$\\
				\hline
				$0.5$ & $0.95$\\
			\end{tabular}
		\end{ruledtabular}
	\end{table}
For the heating system ($\gamma = 0.05$) the VACF reveals a time interval where the auto-correlation stays negative for finite time which cannot be observed for the cooling system ($\gamma = 10$). However, in Fig. \eqref{Fig:PDF_F_alpha_1_DPD_0.5} we see that the acceleration pdf of the latter differs strongly from the behavior of the heating C-DPD and the standard DPD system. We observe non-Gaussian tails of the pdf for which the correlated noise is responsible. We point out that experimental measurements in \cite{la_porta_fluid_2001} reported that the Lagrangian acceleration pdf in fully developed turbulence follows
\begin{equation}
P(a)=C\exp(-a^2 /((1+|a\beta|^{\gamma})\sigma^2)).
\end{equation}
The fit of the realization for the cooling system with the latter function is in surprisingly good agreement with the experiment. The corresponding parameters $C$, $\alpha$, $\gamma$ and $\sigma$ are given in Fig. \eqref{Fig:PDF_F_alpha_1_DPD_0.5}. %The energy spectrum of the cooling system is no less interesting, since it exhibits a slope proportional to $k^{-5/3}$ for some finite range of wave number $k$, see Fig. \eqref{Fig:Ek}. Recall that, for turbulent flow, the second similarity hypothesis predicts that the energy-spectrum function is given by
%\begin{equation}
%	E(k) = D\varepsilon^{2/3}k^{-5/3}
%\end{equation}
%in the inertial subrange. Where $\varepsilon$ is the rate of dissipation of turbulent kinetic energy and $D$ is some constant.

%"[1,1,1,1]" "[0.5,0.05,0.158,10]" "[0,1,1,1]"
%(1.0786555275007366, 59.079711547120745, 0.9453472389068538, 0.5)
%(1, ' C_0=', -0.0013565476085667036)
%((1005309,), (1005309,))
%(1.115859378843673, 7.446182543084814, 0.8991880081112105, 0.05)
%(1, ' C_0=', -0.03847506914084082)
%((1040259,), (1040259,))
%momentum.py:66: RuntimeWarning: invalid value encountered in power
%return  a*exp(-(x*b)**c) #a*exp(-(b*t)**c)
%(1.157982092899605, 22.797747264352378, 0.706404241010902, 0.158)
%(1, ' C_0=', 0.03429860223775475)
%((712165,), (712165,))
%(1.2939867714314586, 467.0766647510949, 0.6275682545371848, 10)
%(1, ' C_0=', -0.0010886417751424786)
%fit by :  a*exp(-(x*b)**c)

\section{Summary and discussions}\label{SummaryAndDiscussions}
The dissipative particle dynamics (DPD) model with correlated stochastic force reveals phenomena beyond of what has been observed so far for standard DPD. In the present paper we have formulated a colored DPD system for which a fluctuation dissipation relation is found empirically without explicit memory kernel in the dissipative force \cite{li_incorporation_2015}. Results show that a proper calibration of the coefficients of the different forces results in a system fluctuating around a constant temperature. %Empirical evidence is provided that a FDT holds. 
The system reaches a unique stationary state, e.g. also shown by the fact that the second moment of the velocity distribution remains bounded with time.

Transport properties determined by the velocity autocorrelation function alone do not differ from standard DPD.  Typical trajectories for Levy flights in the configuration space have been observed with the numerical realizations.

The results of the present paper encourage further investigations of C-DPD with a non-zero conservative force. Also we will consider to investigate the formulation and the effect of a memory kernel which is consistent with the random force.

%\begin{figure}[h]
%	\centering
%	\includegraphics[width=0.4\textwidth]{./Ek_alpha_1_DPD_05.png}
	%Green = gamma = 10, alpha 1 / blue = gamma 0.05, alpha 1 /orange = gamma = 0.158, alpha 1
%	\caption{Energy spectrum is represented in log-log scale. The two extrema of the C-DPD simulation are plotted here. In the case where the dissipation is almost overtaking the fluctuations. Fitting the later one in a specific range gives a curve proportional to $k^{-5/3}$ (dashed black line). }\label{Fig:Ek}
%\end{figure}

\section{Acknowledgments}
We would like to acknowledge support by the Deutsche Forschungsgemeinschaft within the
Priority Programme Turbulent Superstructures under Grant no AD 186/30.

\newpage
\bibliographystyle{unsrtnat}
%\begin{thebibliography}{6}
	%\bibitem{Bourguet} Bourguet, R., Braza, M., Dervieux, A.: Reduced-Order Modeling for Unsteady Transonic Flows Around an Airfoil. Phys.~Fluids \textbf{19}, 111701 (2007)
%\end{thebibliography}

\bibliography{Kangaroo2}{}%,dpd_kangarooNotes}{}

\end{document}